\documentclass[final]{elsarticle}

\usepackage{hyperref}
\usepackage{amsmath}
\usepackage[margin=2cm]{geometry}

\journal{Journal of \LaTeX\ Templates}

\begin{document}

\begin{frontmatter}

\title{COVID-19 mortality prediction: A case study for İstanbul}


\author[mymainaddress]{Erkan Yilmaz}
\ead{erkany@ogr.iu.edu.tr}
\author[mysecondaryaddress]{Özgür Ökcü}
\ead{ozgur.okcu@ogr.iu.edu.tr}
\author[mysecondaryaddress]{Ekrem Aydiner\corref{mycorrespondingauthor}}
\ead{ekrem.aydiner@istanbul.edu.tr}

\cortext[mycorrespondingauthor]{Corresponding author}
\address[mymainaddress]{Institute of Graduate Studies in Science, Istanbul University, 34134, İstanbul, Turkey}
\address[mysecondaryaddress]{Department of Physics, Faculty of Science, Istanbul University, İstanbul, 34134, Turkey}

\begin{abstract}
In this paper, we use SEIR equations to make predictions for the number of mortality due to COVID-19 in İstanbul. Using excess mortality method, we find the number of mortality for the previous three waves in 2020 and 2021. We show that the predictions of our model is consistent with number of moralities for each wave. Furthermore, we predict the number of mortality for the second wave of 2021. We also extend our analysis for Germany, Italy and Turkey to compare the  basic reproduction number $R_0$ for Istanbul. Finally, we calculate the number of infected people in Istanbul for herd immunity.
\end{abstract}


\end{frontmatter}


\section{Introduction}

The world has struggled to control the deadly  coronavirus disease 2019 (COVID-19) which is caused by severe acute respiratory syndrome coronavirus 2 (SARS‑CoV‑2). After the initial outbreak in  China, the novel COVID-19 spread across the world in a short time. COVID-19 was declared a pandemic on 11th March 2020 by World Health Organisation (WHO) \cite{who_pan}. In the same day, the first case in Turkey was reported by the Ministry of Health  \cite{TCSB:1}.

In order to struggle with diseases more effectively and detect their characteristics such as symptoms, spread dynamics, transmission, etc., the humans firstly began to record the diseases.  Those records paved the way to fight against the diseases. The scientists have modelled the pandemics to reveal the pandemics characteristics.  Modelling a pandemic is very important since it may provides how to control the disease and how to reduce the mortality. For example, London mortality statistics were published weekly  in Bills of Mortality from the 17th century to 1830s \cite{heitman}. In 1662, Graunt studied mortality rates and causes in his book Natural and Political Observations Made Upon the Bills of Mortality \cite{david}. 

The first mathematical model on the characteristic behaviours of contagious diseases  was proposed by  Bernoulli in 1766 \cite{Bernoulli}. He investigated the smallpox disease and variolation method which aimed to immunize people against smallpox with the material taken from patients.  Bernoulli showed that variolation was necessary in his mathematical model \cite{Bernoulli}. In 1906, Hamer analysed the measles epidemic and proposed a discrete time model  \cite{hamer}. This model is important since it shows that the new cases of an epidemic depend on the numbers of patients and  individuals susceptible to disease. In 1911,  Ross proposed a differential equation model related to the number of cases and control of malaria \cite{Bacaer}. In 1927, Kermack and McKendrick published their paper entitled A Contribution to the Mathematical Theory of Epidemics \cite{Kermack:1927}. In their paper, they considered an interactive model for an isolated society. They proposed SIR model. In Refs.\cite{Kermack:1932, Kermack:1933}, they made the model more useful by adding population dynamics and developed SIR model.

Building a mathematical model provides predictions for pandemics, effective control and political strategies  \cite{anderson,scarpno,chinazzi,rudige,bilge}. It also provides accurate future predictions and effective struggle against the diseases  \cite{wangping,kucharski,liq}. Differential equations can be used to predict the time evolution of the pandemic. Although the mathematical models are useful to predict the characteristics of a disease, they have some difficulties. It is not easy to determine the model parameters. Besides, the factors such as mutations, health infrastructure, disasters, etc. affect the evolution of a disease. Therefore, the predictions from differential equations may become harder.  

In order to determine the COVID-19 characteristics, the COVID-19 has been widely studied in the literature for two years \cite{wangping,danon,fanelli,roosa,li,Peng,mangoni}. In order to predict the trend of COVID-19 in Italy, Wangping et. al. used the extended SIR model \cite{wangping}. They also estimated the basic reproductive number $R_0$ from Markov Chain Monte Carlo methods. Danon et. al. adapted an existing national-scale metapopulation model to get the spread of CoVID-19 in England and Wales \cite{danon}. They found that the size and spread rate of epidemic highly depend on the seasonal transmission. Fanalli and Piazza analyzed the temporal dynamics of the COVID-19 outbreak in China, Italy and France \cite{fanelli}. Roosa et. al. \cite{roosa} reported short-term estimation of the cumulative number of reported cases of the COVID-19 epidemic in Hubei province and other provinces in China as of February 9, 2020. Using the data until February 9, 2020, they also estimated the number of reported cases between 37,415 and 38,028 in Hubei Province and 11,588–13,499 in other provinces by February 24, 2020. In Ref. \cite{li}, Li et. al. studied the transmission process of the COVID-19 via based on the official data. Based on Ref. \cite{Peng}, Mangoni and Pistilli \cite{mangoni} developed a generalised SEIR model to estimate the dynamics of COVID-19.

Up to now, Covid-19 has the best statistics. However, recording exact and reliable data include serious difficulties. For example, PCR tests expected success rate is very low. In addition to medical issues, political and economic issues affect the reliability of these data. On the other hand, mortality records of a country are more reliable, and excess mortality data are independent from political, economic and medical issues. 

As before mentioned, the differential models provides to determine the behaviours of an epidemic despite its disadvantages. SEIR is the one of these models \cite{anderson}. In thse paper, we use the SEIR model to predict the COVID-19 waves for Istanbul. To the best of our knowledge, we calculate the parameters of SEIR model from excess mortality methods for the first time. Therefore, we believe our results are reliable since they are based on excess mortality. 

The paper is organised as follows: In Section 2, we  review the SIR and SEIR models. We present the details on determining the parameters $\beta$, $\eta$ and $\epsilon$ of SEIR model, excess mortality methods, and time series for COVID-19. In Section 3, we determine the time series from excess mortality methods. We also find the SEIR model parameters. We calculate the basic reproduction number $R_0$ for each wave. We make predictions for the second wave of 2021. In order to compare to value of $R_0$ for Istanbul, we also computed $R_0$ for Germany, Italy and Turkey. In Section 4, we finally discuss our results and estimate the number of infected people for herd immunity. 

\section{Model}

\subsection{Mathematical models of the pandemic}

In this study, we use the  SEIR model \cite{anderson} which is the modification of SIR model. Just like SIR model, SEIR is also a set of coupled differential equations. It is possible to construct many disease spread models from SIR and modified SIR models\cite{Kermack:1927,Kermack:1932,Kermack:1933,bilge,wangping,Peng,mangoni}. Recently, much attention has been paid to these models due to COVID-19. Because these models provide to estimate numbers of dead, recovered and infected individuals for local and global regions during a pandemic. These models completely depend on the parameters and give the reliable results.  However, contagious diseases may regard as  catastrophic events. Because stochastic factors such as vaccine, mutations, curfew or quarantine precautions may make the process more chaotic.  Therefore, it is difficult to make the accurate predictions. On the other hand, it is not impossible to make accurate predictions for a short term since all stochastic effects can be expressed in macroscopic forms in the limit cases.

SIR equations are given by \cite{Kermack:1927}

\begin{eqnarray}
	\frac{dS}{dt}  &=& -\beta S I \nonumber, \\
	\frac{dI}{dt} &=&  \beta S I - \eta I \nonumber, \\
	\frac{dR}{dt} &=& \eta I, 	
	\label{equ:SIR}
\end{eqnarray}
where $S$, $I$ and $R$ correspond to susceptible, infected and recovered individuals, respectively. Total probability is $S+I+R=1$. Let us explain $S$, $I$ and $R$, respectively. $S$ represents the individuals that may have tendency to get a disease. $I$ represents the individuals that are infected and spread a disease. $R$ represents the individuals that have not got any possibilities to get a disease. Individuals which is recovered or died from a disease can be considered in $R$. The parameter $\beta$ corresponds to disease spread speed while the parameter $\eta$ is related with illness duration. We define the parameter basic reproduction number as $R_0=\frac{\beta}{\eta}$. It is important for the disease spread dynamics. 

There are some disadvantages of SIR model. Getting a disease and contagiousness are considered as mono-type. In fact, these factors vary from individual to individual. The spread rate of an epidemic may depend on seasons. A disease may have different effects from human to human since humans have different genetics.

A more realistic model is the modification of SIR model, i.e., SEIR model \cite{anderson}. In SEIR model, there is an extra parameter $E$ which represents the number of individuals exposed to COVID-19. When people exposed to coronavirus, they do not immediately become ill. There is a time period between getting the virus and transmitting to humans. This period is called incubation period. The incubation period changes the date and value of peak. SEIR equations are given by \cite{anderson}

\begin{eqnarray}
	\frac{dS}{dt} &=& -\beta S I \nonumber \\
	\frac{dE}{dt} &=& \beta S I- \epsilon E  \nonumber \\
	\frac{dI}{dt} &=& \epsilon E - \eta I  \nonumber \\
	\frac{dR}{dt} &=& \eta I 
	\label{equ:SEIR}
\end{eqnarray}
where the parameter $\epsilon$ is related with the incubation period of epidemic.

\subsection{Determining the model parameters}

Excess mortality determines the number of deaths from all reported number of deaths during a crisis and expected deaths for the same time in the absence of the crisis  \cite{ritche,checchi}. Excess mortality method was used to determine the moralities of  the Great Plague of London in 1665 \cite{boka}, influenza epidemic in London in 1875 \cite{farr,langmuir}, influenza pandemics of 1918, 1957, 1968, 2009 \cite{murray, vibound} , as well as seasonal influenza epidemics \cite{housworth}. Excess mortality is given by \cite{ritche}
\begin{equation}
	Excess\, Mortality = Reported\, Deaths - Expected\, Deaths .
\end{equation}

\begin{figure} [h]
	\centering
	\includegraphics[width=12cm]{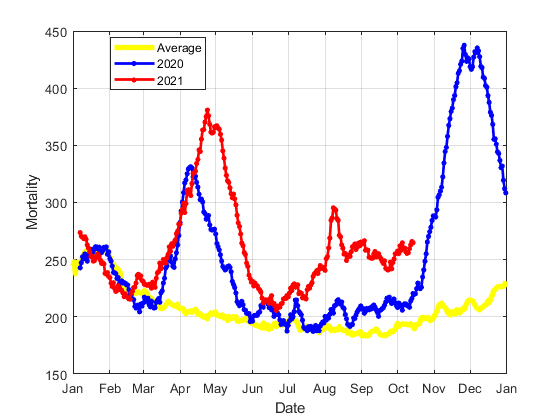}	
	\caption{Numbers of reported death for 2015-2019 (yellow), 2020 (blue) and 2021 (red).}
	\label{7years}
\end{figure}

\begin{figure} [h]
	\centering
	\includegraphics[width=12cm]{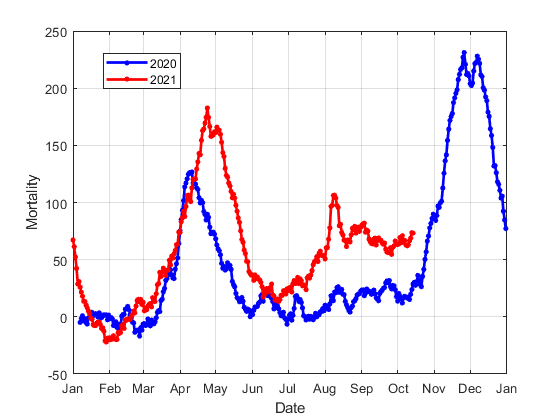}
	
	\caption{Using excess mortality method, numbers of death due to the COVID-19 in 2020 (blue) and 2021 (red).}
	\label{mortality}
\end{figure}

\section{Istanbul Covid-19 Waves}
\subsection{The first wave of 2020}

This method is more reliable to detect the number of deaths from COVID-19 \cite{krelle}. Using excess mortality method, WHO has obtained the number of deaths from seasonal influenza, pandemics and other public health threats since 2009 \cite{who:2020}.

We take numbers of  daily deaths in İstanbul from 2015 to 2019 \cite{edevlet} to predict the expected number of death in 2020-2021. In the calculation of expected deaths, $40\%$ deaths in 2019, $30\%$ deaths in 2018, $20\%$ deaths in 2017, $5\%$ deaths in 2016 and $5\%$ deaths in 2015 were taken. In order to make our time series more reliable, we consider the following average over seven days 
\begin{equation}
	d = \frac{d_i+d_{i-1}+d_{i-2}+d_{i-3}+d_{i-4}+d_{i-5}+d_{i-6}}{7}
	\label{equ:s}
\end{equation}

In Fig. (\ref{7years}), we present the reported numbers of deaths in  2015-2019, 2020, 2021. Yellow line represents the numbers of average mortality between 2015 and 2019. Note that value of yellow line is between 200 and 250. Furthermore, numbers of reported death for 2020 (blue line) and 2021 (red line) reach bigger values than yellow line's. It is clearly obvious that numbers of mortality dramatically increased during COVID-19 pandemic. In Fig. (\ref{mortality}), using excess mortality method for the data between 2015-2019, we show numbers of death in 2020 and 2021 due to COVID-19. In this figure, it is clear that there are two COVID-19 waves for each year.  

The first case in Turkey was reported on March 2020 and COVID-19 quickly spread across the country. The first wave in Istanbul occurred in the spring of 2020. Pandemic started on 15th March 2020 and 1653 people died from COVID-19 until pandemic's first peak 10th April 2020. Then it started to decrease. During the decrease of the first wave, 2798 people died from COVID-19.

\begin{figure} [!ht]
	\centering
	\includegraphics[width=6cm]{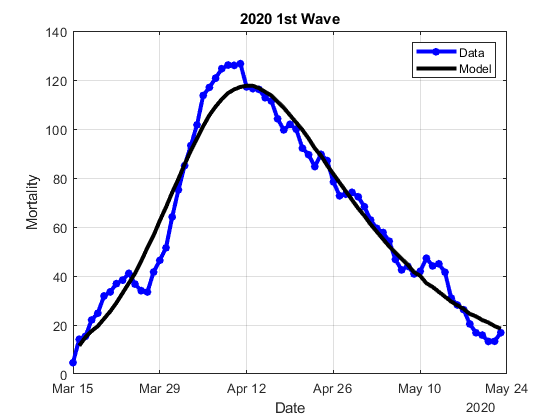}
	\includegraphics[width=6cm]{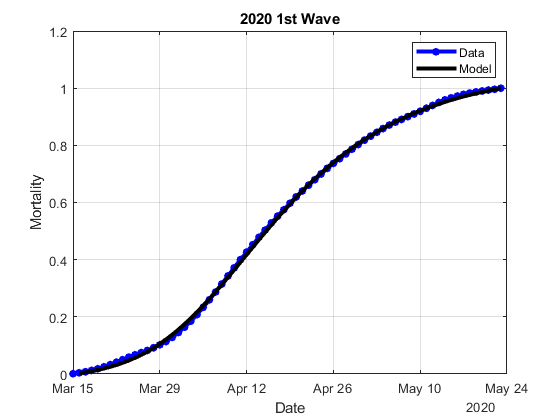}
	\caption{Mortality numbers of the first wave in 2020 (a)Daily (b)Total}
	\label{2020_1st}
\end{figure}

The first wave in Istanbul reached the peak in 26 days, then it decreased in 43 days. The total time of the first wave is 69 days. 4451 people died from COVID-19 during the first wave. In Fig. (\ref{2020_1st}), we show daily and total moralities. It can be seen that positive slope of graphics is bigger than negative slope. This implies that the value of $R_{0}$ is high.

\begin{figure} [!h]
	\centering
	\includegraphics[width=6cm]{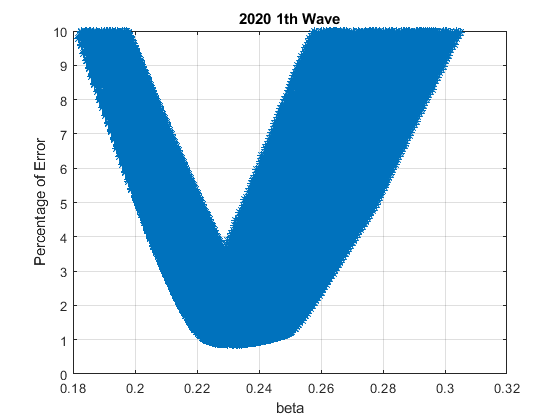}
	\includegraphics[width=6cm]{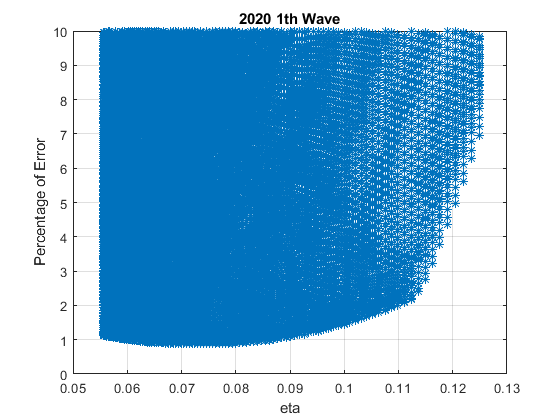}
	\caption{Percentages of error for the first wave in 2020 (a) The parameter $\beta$ (b) The parameter $\eta$ for the first wave.}
	\label{beta_20201}
\end{figure}

In order to detect the parameters $\beta$, $\eta$ and $\epsilon$ of the first wave, we must obtain the numerical solutions. First of all, we define the SEIR model equations and data in matlab program. We obtain the first wave parameter within 1 or 2 percentage of error. The parameters and their percentages of error can be found in Fig.( \ref{beta_20201}). Comparing SEIR model with data gives more than 10,000 results within 5 percentage of error. Best ten results are listed in Table  \ref{tab:para2020_1}. The results in Table \ref{tab:para2020_1} are obtained less than 1 percentage of error.  We use the fourth result in Table \ref{tab:para2020_1} to produce Fig. (\ref{2020_1st}) since it is close to average value.

\begin{table}[ht!]
	\centering
	\caption{The First Wave Parameters of 2020}
	\label{tab:para2020_1}
	\begin{tabular}{|l|l|l|l|l|}
			\hline
		$R_0=\frac{\beta}{\eta}$   & $\beta$          & $\eta$     &$\epsilon$ & Error $\%$  \\\hline
	3.19	&0.232846715	&0.072992701	&3	&0.867405 \\ \hline
	3.11	&0.232089552	&0.074626866	&3	&0.868163 \\ \hline
	3.22	&0.23	        &0.071428571	&4	&0.868588 \\ \hline
	3.16	&0.232352941	&0.073529412	&3	&0.868712 \\ \hline
	3.22	&0.233333333	&0.072463768	&3	&0.868797 \\ \hline
	3.14	&0.232592593	&0.074074074	&3	&0.86956 \\ \hline
	3.14	&0.22919708	    &0.072992701	&4	&0.869781 \\ \hline
	3.25	&0.230496454	&0.070921986	&4	&0.869876 \\ \hline
	3.08	&0.231578947	&0.07518797	    &3	&0.870172 \\ \hline
	3.17	&0.229710145	&0.072463768	&4	&0.870274 \\ \hline
		\end{tabular}
\end{table}

\subsection{The Second Wave of 2020}

The second wave started on 22nd October 2020. It peaked on 26th November 2020, and ended on 14th January 2021. Mortality increased for 35 days, and decreased for 49 days. The second wave lasted 84 days. 4754 people died from COVID-19 until the peak. During the decrease of the second wave, 6433 people died. 11187 people died from COVID-19 during the second wave. 

Comparing the first and second waves, we can easily see differences between them. The second wave is more symmetrical than the first wave. It means that the value of $R_{0}$ is more smaller than the first wave's. The number of mortality for the second wave is more higher than the first wave's. The mutation of the virus or seasonal effects may have caused more deaths.

\begin{figure} [ht!]
	\centering
	\includegraphics[width=6cm]{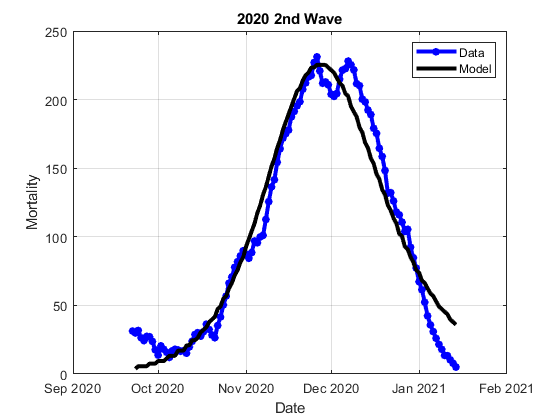}
	\includegraphics[width=6cm]{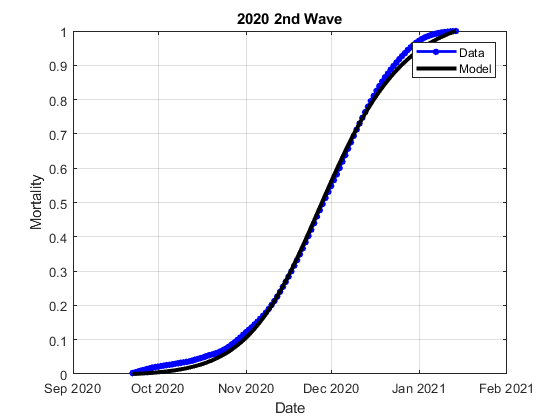}
	\caption{Mortality numbers of the second wave in 2020 (a)Daily (b)Total}
	\label{2020_2nd}
\end{figure}

\begin{figure} [!ht]
	\centering
	\includegraphics[width=6cm]{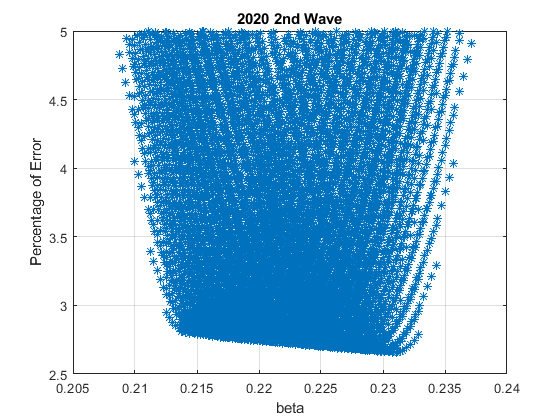}
	\includegraphics[width=6cm]{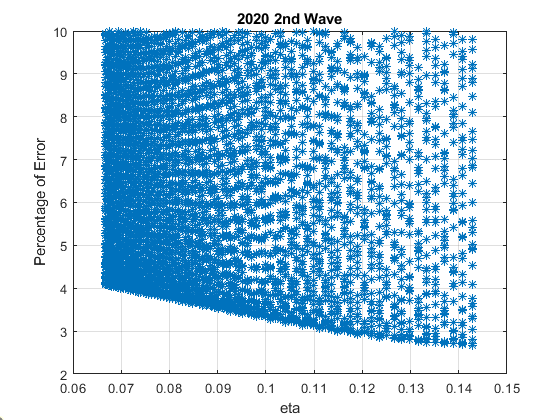}
	\caption{Percentages of error for the second wave in 2020 (a) The parameter $\beta$ (b) The parameter $\eta$ for the second wave.}
	\label{beta_20202}
\end{figure}

The numbers of mortality for the second wave  are given in Fig. (\ref{2020_2nd}). We calculate the parameters for the second wave. We found the percentages of error for the parameters $\beta$ and $\eta$ in Fig. (\ref{beta_20202}). In Fig. (\ref{beta_20202}.a), one can see that the parameter $\beta$ has a low percentage of error between 0.215 and 0.235. Its percentage of error has the lowest value near 0.23. In Fig. (\ref{beta_20202}.b), the parameter $\eta$ has a low percentage of error between 0.13 and 0.15. It has the lowest percentage of error near 0.14. Ten results which give the lowest percentage of error are given in Table (\ref{tab:para2020_2}). We use the first result in Table (\ref{tab:para2020_2}) to produce the Fig. \ref{2020_2nd}.

\begin{table}[!ht]
	\centering
	\caption{The Second Wave Parameter of 2020}
	\label{tab:para2020_2}
	\begin{tabular}{|l|l|l|l|l|}
		\hline
				$R_0 =\frac{\beta}{\eta}$   & $\beta$          & $\eta$     & $\epsilon$ & Error $\%$  \\\hline
	1.62	&0.231098431	&0.142653352	&3	    &2.655150333 \\\hline
	1.62	&0.230769231	&0.142450142	&3		& 2.656684896 \\\hline
	1.62	&0.231428571	&0.142857143	&3		&2.661003256 \\\hline
	1.62	&0.229787234	&0.141843972	&3		&2.661754709 \\\hline
	1.63	&0.230878187	&0.141643059	&3		&2.662501288 \\\hline
	1.62	&0.230113636	&0.142045455	&3		&2.662901328 \\\hline
	1.63	&0.229901269	&0.141043724	&3		&2.663813534 \\\hline
	1.63	&0.231205674	&0.141843972	&3		&2.664873865 \\\hline
	1.62	&0.230440967	&0.142247511	&3		&2.665545267 \\\hline
	1.63	&0.229577465	&0.14084507	    &3		&2.666175529 \\\hline	
	\end{tabular}
\end{table}

\subsection{The first wave of 2021}

The first wave of 2021 started on 11th March 2021. It peaked on 20th April 2021, and ended on 8th June 2021. Mortality increased for 44 days, and decreased for 45 days. 3886 people died until the peak. During the decrease of the wave, 4422 people died. 8308 people died during the first wave of 2021. 

The numbers of mortality are more higher than the numbers of mortality in the first wave of 2020. The first wave of 2021 is more symmetrical than the first wave of 2020.

\begin{figure} [ht!]
	\centering
	\includegraphics[width=6cm]{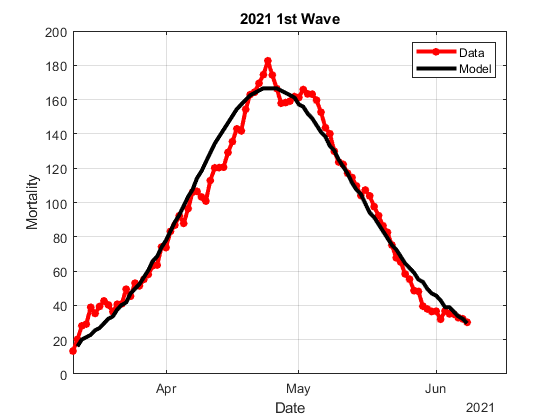}
	\includegraphics[width=6cm]{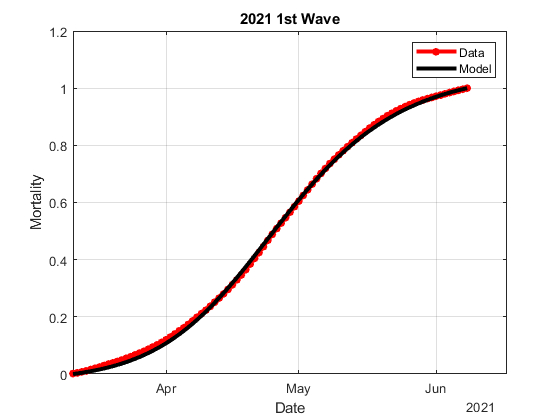}
	\caption{Mortality numbers of the first wave in 2021 (a)Daily (b)Total}
	\label{2021_1st}
\end{figure}

\begin{figure} [ht!]
	\centering
	\includegraphics[width=6cm]{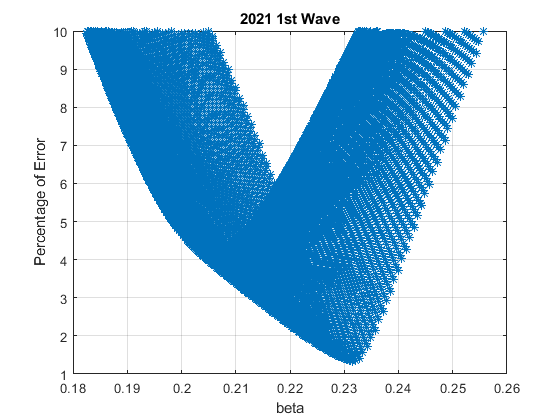}
	\includegraphics[width=6cm]{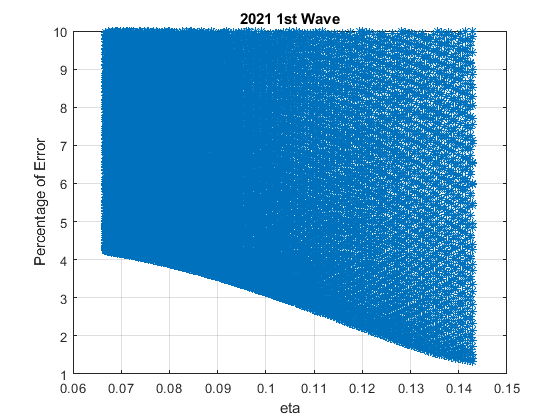}
	\caption{Percentages of error for the first wave in 2021 (a) The parameter $\beta$ (b) The parameter $\eta$ for the first wave.}
	\label{beta_20211}
\end{figure}

The numbers of mortality are given in Fig. (\ref{2021_1st}). The percentages of error for parameters $\beta$ and $\eta$ are given in Fig. (\ref{beta_20211}). In Fig. (\ref{beta_20211}.a), it can be seen that the parameter $\beta$ has the low percentage of error between 0.21 and 0.24. Its percentage of error is the lowest near 0.23.  In Fig (\ref{beta_20211}.b), one can see that the parameter $\eta$ has a low percentage of error between 0.13 and 0.15. In Table \ref{tab:para2021_1}, its percentage of error is the lowest near 0.14. We list the parameters which have the lowest error of percentage. We use the second result of Table (\ref{tab:para2021_1}) to produce the Fig. (\ref{2021_1st}). 

\begin{table}[!ht]
	\centering
	\caption{The First Wave Parameters of 2021}
	\label{tab:para2021_1}
	\begin{tabular}{|l|l|l|l|l|}
		\hline
		
		$R_0 =\frac{\beta}{\eta}$   & $\beta$          & $\eta$     & $\epsilon$ & Error $\%$  \\\hline
		
		1.61	&0.23	        &0.142857143	&3.56	&1.324750255	 \\\hline
		1.61	&0.229344729	&0.142450142	&3.75	&1.328061544	\\\hline
		1.6	    &0.228571429	&0.142857143	&4.19	&1.330648971	\\\hline
		1.61	&0.228693182	&0.142045455	&3.96	&1.331516693	\\\hline
		1.6	    &0.228571429	&0.142857143	&4.12	&1.331578669	\\\hline
		1.61	&0.228045326	&0.141643059	&4.19	&1.335107124	\\\hline
		1.6	    &0.227920228	&0.142450142	&4.44	&1.335252725	\\\hline
		1.62	&0.229461756	&0.141643059	&3.53	&1.338365318	\\\hline
		1.61	&0.22740113	    &0.141242938	&4.53	&1.340113226	\\\hline
		1.62	&0.228813559	&0.141242938	&3.73	&1.340357901	\\\hline

	\end{tabular}
\end{table}

\subsection{The predictions on the second wave of 2021}

Up to now, we have modelled the COVID-19 by using SEIR and found the model parameters $\beta$, $\eta$ and $\epsilon$. These parameters determine the behaviours of the pandemic.  The parameter $\epsilon$ does not effect the trend of epidemic but changes the peak date of epidemic. Now, using the Tables \ref{tab:para2020_1}, \ref{tab:para2020_2} and \ref{tab:para2021_1}, we will predict the numbers of deaths for the second wave of 2021. In Table \ref{tab:predicbefore}, we give the results of each wave for lowest error of percentage.

\begin{figure} [ht!]
	\centering
	\includegraphics[width=12cm]{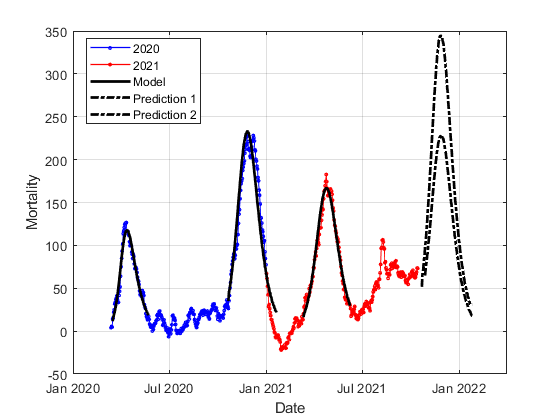}
	
	\caption{Numbers of mortality for last three COVID-19 waves. Dashed black lines correspond to predictions for the current COVID-19 wave.}
	\label{prediction}
\end{figure}

\begin{table}[!ht]
	\centering
	\caption{The averages of ten results which have the lowest percentages of error.}
	\label{tab:predicbefore}
	\begin{tabular}{|l|l|l|l|}
		\hline
		
		Wave & $R_0 =\frac{\beta}{\eta}$   & $\beta$          & $\eta$       \\\hline
		
		First wave of 2020	&3.168	&0.231419776	&0.073068182	 \\ \hline
		Second wave of 2020	&1.624	&0.230520067	&0.14194734	      \\ \hline 
		First wave of 2021	&1.609	&0.228682277	&0.142128916	  \\ \hline
	\end{tabular}
\end{table}
As it can be seen in Table \ref{tab:predicbefore}, the values of $R_0$ are 3.168 for the first wave of 2020,  1.624 for the second wave of 2020 and 1.609 for the first wave of 2021. The second and third values of $R_0$ are lower than the first value. It means that the spread speed decreased. Because taking precautions against COVID-19 is very effective. We assume that the parameters value of the second wave of 2021 is nearly close to last two waves. These parameters may decrease due to vaccination or may increase due to mutation. Therefore, we estimate lower and upper bounds for the new wave. In Fig. (\ref{prediction}), we present our predictions for the second wave of 2021. In the figure, we show the numbers of mortality for 2020 and 2021. It can easily be seen that our model (black line) is fit with 2020 (blue line) and 2021 (red line) data. We present our predictions with black dashed line. 

\subsection{Calculation of $R_0$ for Turkey, Germany and Italy  }

Up to now, we have considered $R_0$ for Istanbul. Taking precautions cause to decrease the number of cases. Therefore, the parameter $R_0$ has tendency to decrease. In this section, we want to study the behaviour of $R_0$ for Germany, Italy and Turkey. The value of basic production number $R_{0}$ for the first wave of 2020 is more bigger than the value of $R_0$ for the other waves. In order to test it, we found the values of $R_0$ for the numbers of case for Germany, Italy and Turkey.

We take the data from the web site of  Republic of Turkey Ministry of Health \cite{TCSB:1}.  The first wave of 2020 and the first wave of 2021 are estimated for the numbers of case. However, we take into account the numbers of mortality for the second wave of 2020 due to the change of the COVID-19 new case announcement system. In Table \ref{tab:turkeywave}, one can see the values of $R_0$ decreasing for Turkey. 

\begin{table}[!ht]
	\centering
	\caption{$R_0$ for the COVID-19 Waves in Turkey}
	\label{tab:turkeywave}
	\begin{tabular}{|l|l|l|}
		\hline
		
		Year & Wave       & $R_0 =\frac{\beta}{\eta}$    \\\hline
		2020 	& First 	& 4.2 - 4.7  	\\ \hline
		2020 	& Second 	& 1.3 - 1.47 	\\ \hline
		2021 	& First 	& 1.75 - 2.1 	\\ \hline

\end{tabular}
\end{table}

In Table \ref{tab:DEUwave}, using the data in Ref. (\cite{ourWorldData}), we give $R_{0}$ for Germany. It can be seen that $R_0$ gets the values between 5 and 9 for the first wave of 2020. If such huge values were permanent, all the people in the world would nearly have COVID-19. For the second wave of 2020, $R_0$ has the values between $1.3$ and $2.1$. Finally, for the first wave of 2021, $R_0$ has the values between $1.4$ and $1.8$. 

\begin{table}[!ht]
	\centering
	\caption{$R_0$ for the COVID-19 Waves in Germany}
	\label{tab:DEUwave}
	\begin{tabular}{|l|l|l|}
		\hline
		
		Year & Wave       & $R_0 =\frac{\beta}{\eta}$    \\\hline
		2020 	& First 	& 5.0 - 9.0 	\\ \hline
		2020 	& Second	& 1.3 - 2.1	\\ \hline
		2021 	& First	    & 1.4 - 1.8	\\ \hline
		
	\end{tabular}
\end{table}

For Italy, using the data in Ref. (\cite{ourWorldData}),  one can see the values of  $R_0$ in Table \ref{tab:itawave}. The values of $R_0$ are between 4 and 5 for the first wave of 2020. The values of $R_0$ are between $3.5$ and 5 for the second wave of 2020.    For the first wave of 2021, $R_0$ has the values between $1.3$ and $1.9$.

\begin{table}[!ht]
	\centering
	\caption{$R_0$ for the COVID-19 Waves in Italy}
	\label{tab:itawave}
	\begin{tabular}{|l|l|l|}
		\hline
		Year & Wave       & $R_0 =\frac{\beta}{\eta}$    \\\hline
		2020 	& First 	& 4.0 - 5.0 	\\ \hline
		2020 	& Second 	& 3.5 - 5.0		\\ \hline
		2021 	& First 	& 1.3 - 1.9		\\ \hline
	\end{tabular}
\end{table}

\begin{figure} [ht!]
	\centering
	\includegraphics[width=6cm]{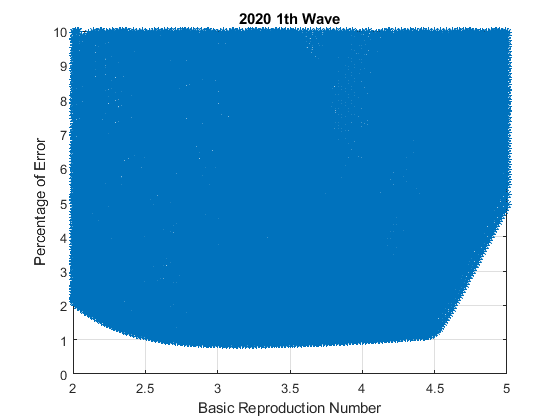}
	\includegraphics[width=6cm]{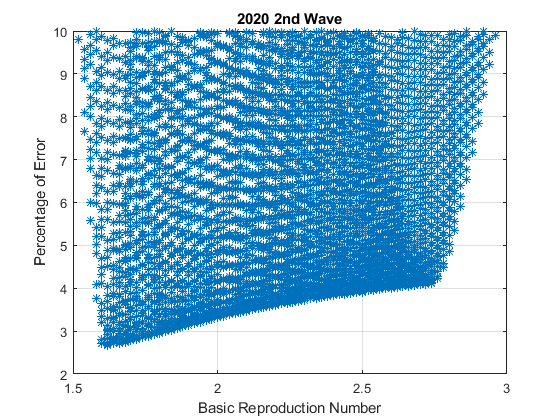}
	\includegraphics[width=6cm]{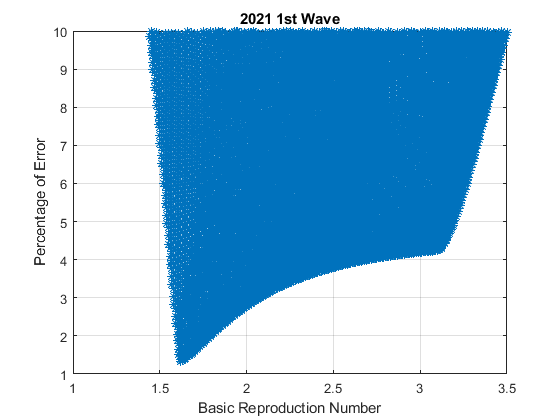}
	\caption{$R_{0}$ for Istanbul (a) The first wave of 2020 (b) The second wave of 2020 (c) The first wave of 2021}
	\label{BRPOfIst}
\end{figure}

Our analysis show that $R_0$ may have the tendency to decrease in the presence of precautions. 

\section{Conclusions and discussions}
In this paper, using the reported numbers of mortality for 2014-2021, we have determined the numbers of COVID-19 mortality in Istanbul via excess mortality method. Our analysis shows that there are two deadly pandemic waves for a year. We have modelled  each wave  via SEIR equations. We have calculated the parameters $\beta$, $\eta$, $\epsilon$ and $R_0$. Our model is  compatible with the previous three waves. Then, we have estimated the second wave of 2021. Interestingly, we have found that $R_0$ may have tendency to decrease for precautions. In order to confirm the behaviour of $R_0$, we have extended our analysis for Germany, Italy and Turkey. It seems $R_0$ may decrease for precautions. \footnote{Mutation may change this case. Omicron is one of the recent mutation, which is very contagious. In this paper, we do not take into account any mutations. We will left this case for future study.}

\begin{figure}[ht]
		\centering
	\includegraphics[width=14cm]{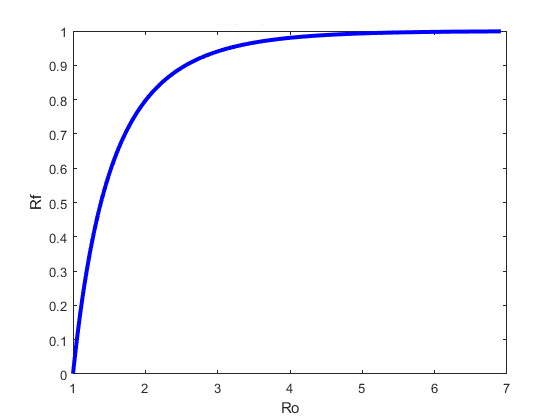}
	\caption{The behavior of $R_f$ with respect to $R_0$.}
	\label{Ro}
\end{figure}

Before finishing the paper, we give some comments on the number of infected people to gain herd immunity in Istanbul. Solving Eq. (\ref{equ:SIR}), one can determine the how many individuals get infected with COVID-19 until the end of the pandemic.\footnote{In fact, we can use SEIR equations, but the result does not change. The extra parameter $\epsilon$ only changes the peak date of an epidemic.} Using 
\begin{equation}
	S+I+R=1
	\label{equ:esit1}
\end{equation}
with the initial conditions $S(0)=1$ and $R(0)=0$, one finds
\begin{equation}
	S(t)=e^{-R_0R(t) }
	\label{degisken1}.
\end{equation}
At the end of COVID-19, the value of $I$ is zero. Therefore, employing Eq. (\ref{degisken1}) in Eq. (\ref{equ:esit1}), we obtain
\begin{equation}
	R_f + e^{-R_0 R_f} =1,
	\label{equ:rfinal}
\end{equation}
where $R_f$ is the number of recovered people at the end of COVID-19. We show numerical solutions of this equation in Fig. (\ref{Ro}).
In Fig. (\ref{Ro}), one can see that all people nearly get infected for $R_0=5$. For example, varying the value of $R_0$ from 7 to 5 does not effect the number of infected people. However, varying the value of $R_0$ from 5 to 3 effects the number of infected people.

In Fig. (\ref{BRPOfIst}), we show the $R_0$ for the previous three waves in Istanbul. Using the data in Fig. (\ref{BRPOfIst}), we give the percentage of infected people for herd immunity in Table (\ref{tab:herdUm}). One can easily see that $R_f$ depends on $R_0$. For the first wave of 2020, $89.3\%$-$98.9\%$ of people in Istanbul should get infected to gain herd immunity. The rates for two next waves decrease. Therefore, number of infected people for herd immunity decreases. 

\begin{table}[]
    \centering
        \caption{$R_{0}$ and $R_{f}$ values for different waves in the last two years}
\begin{tabular}{|c|c|c|}
\hline 
Wave & $R_{0}$ & $R_{f}$\tabularnewline
\hline 
\hline 
First wave of 2020 & $2.5$ & $89.3\%$\tabularnewline
\cline{2-3} 
 & $4.5$ & $98.9\%$\tabularnewline
\hline 
Second wave of 2020 & $1.6$ & $64.2\%$\tabularnewline
\cline{2-3} 
 & $1.65$ & $66.8\%$\tabularnewline
\hline 
First wave of 2021 & $1.6$ & $64.2\%$\tabularnewline
\cline{2-3} 
 & $1.63$ & $65.8\%$\tabularnewline
\hline 
\end{tabular}
    \label{tab:herdUm}
\end{table}
We see that the first and second main waves of 2020 and 2021 show seasonal dependence. It can be concluded that the first main mortality wave appears in the spring period which correspond to the increasing of the human metabolic activity. Similarly, the second main mortality wave appears in between December and February. This period  corresponds to the decreasing of the human metabolic activity and increasing air pollution. 

The results obtained in this type of study may contribute to the development strategies of the control, quarantine, physical distance, vaccination and determination of health load. On the other hand, the presented model can be used to make predictions not only local region but also country scale.

\section*{Acknowledgments}

Authors would like to thank Professor Ayşe H. Bilge (Kadir Has University) for useful discussions.






\end{document}